\documentclass[a4paper,11pt]{article}

\usepackage{jheppub} 

\begin{document}

  \title{Tensor renormalization group approach to four-dimensional complex $\phi^4$ theory at finite density}

  \author[a]{Shinichiro Akiyama,}
  \affiliation[a]{Graduate School of Pure and Applied Sciences, University of Tsukuba, Tsukuba, Ibaraki
    305-8571, Japan}

  \author[b,c]{Daisuke Kadoh,}
  \affiliation[b]{Physics Division, National Center for Theoretical Sciences, National Tsing-Hua University,\\ Hsinchu, 30013, Taiwan}
  \affiliation[c]{Research and Educational Center for Natural Sciences, Keio University, \\ Yokohama 223-8521, Japan}

  \author[d]{Yoshinobu Kuramashi,}
  \affiliation[d]{Center for Computational Sciences, University of Tsukuba, Tsukuba, Ibaraki
    305-8577, Japan}

  \author[e]{Takumi Yamashita,}
  \affiliation[e]{Faculty of Engineering, Information and Systems, University of Tsukuba, Tsukuba, Ibaraki
    305-8573, Japan}

  \author[d]{Yusuke Yoshimura}

\abstract{

Tensor network is an attractive approach to field theory with 
negative sign problem. 
The complex $\phi^4$ theory at finite density is a test bed for numerical algorithms to verify their effectiveness.
The model shows a characteristic feature called the Silver Blaze phenomenon 
associated with the sign problem 
in the large volume limit at low temperature. 
We analyze the four-dimensional model  employing the anisotropic tensor renormalization group algorithm. 
We find a clear signal of the Silver Blaze phenomenon on a large volume of $V=1024^4$, 
which implies that the tensor network approach is effective even for four-dimensional field theory beyond two dimensions. 
}

\date{\today}

\preprint{UTHEP-747, UTCCS-P-131, NCTS-CMT/2001}

\maketitle

\section{Introduction}
\label{sec:intro}

The tensor renormalization group (TRG), which was originally proposed by Levin and Nave to study two-dimensional ($2d$) classical spin models \cite{Levin:2006jai}\footnote{In this paper the TRG method or the TRG approach refers to not only the original numerical algorithm proposed by Levin and Nave but also its improved versions \cite{PhysRevB.86.045139,Adachi:2019paf,Kadoh:2019kqk}.}, is a deterministic method to evaluate the partition function.  This method has several superior features over the Monte Carlo method. 
It does not suffer from the sign problem and its computational cost depends on the system size only logarithmically. 
In addition, 
the partition function or the path-integral itself can be directly evaluated and Grassmann variables are also directly manipulated. 
In past several years these features have been confirmed by exploratory numerical studies of the $2d$ 
quantum field theories \cite{Shimizu:2012zza,Shimizu:2014uva,Shimizu:2014fsa,Shimizu:2017onf,Takeda:2014vwa,Kawauchi:2016xng,Kadoh:2018hqq,Sakai:2017jwp,Yoshimura:2017jpk,Unmuth-Yockey:2018ugm,Kadoh:2018tis,Butt:2019uul,Kadoh:2019ube,Kuramashi:2019cgs}. 
A next research direction could be an application of the TRG method to the $4d$ 
quantum field theories following the first study 
of a $4d$ spin model 
\cite{Akiyama:2019xzy,Akiyama:2019chk}. 
  
In this paper we choose the $4d$ complex $\phi^4$ theory at finite chemical potential, which is a typical system 
with the sign problem. 
This model has a characteristic feature called the Silver Blaze phenomenon, which has been studied by various 
methods intended to overcome or tame the sign problem, such as the complex Langevin approach \cite{Aarts:2008wh}, the thimble method \cite{Cristoforetti:2013wha,Fujii:2013sra,Mori:2017nwj}, and the worldline representation \cite{Gattringer:2012df,Orasch:2017niz}. 
Since the $2d$ case was investigated
using the TRG method \cite{Kadoh:2019ube}, we already have a knowledge of 
how to treat the complex scalar field effectively with it. 
The most important change from the $2d$ case 
to the $4d$ one is a choice of the algorithm. The original TRG algorithm proposed by Levin and Nave, which was 
employed in the $2d$ case~\cite{Kadoh:2019ube}, is not applicable to higher dimensional ($\ge 3d$) models. 
Although this defect was overcome by the higher order TRG (HOTRG) algorithm \cite{PhysRevB.86.045139}, 
its computational cost is $O(D^{4d-1})$ with $D$ the bond dimension of the tensors, which is rather expensive 
for higher dimensional models. Recently, a new algorithm called Anisotropic TRG (ATRG) was proposed to reduce 
the computational cost to be $O(D^{2d+1})$ \cite{Adachi:2019paf}. The effectiveness of this algorithm is already 
tested using the $4d$ Ising model \cite{Akiyama:2019chk}. We employ the ATRG algorithm to investigate the $4d$ complex $\phi^4$ theory at finite chemical potential.    
  
This paper is organized as follows. In Sec.~\ref{sec:method} we explain the details of the computational
 method for the $4d$ complex $\phi^4$ theory at finite chemical potential with the ATRG algorithm. 
 Numerical results including the Silver Blaze phenomenon are presented in Sec.~\ref{sec:results}.  Section~\ref{sec:summary} is devoted to summary and outlook.

\section{Method}
\label{sec:method}

\subsection{Tensor network representation}

The complex scalar field theory at finite chemical potential in $4d$ euclidean space 
is given by
\begin{align}
\label{eq:cont_action}
S= \int d^4 x \left\{ |\partial_\nu \phi|^2 + m^2  |\phi|^2 + \lambda |\phi|^4 + \mu ( \partial_4 \phi^* \phi -  \phi^* \partial_4 \phi )
\right\}(x)
\end{align}
with the complex scalar field $\phi(x)$, the bare mass $m$, 
the quartic coupling constant $\lambda>0$ and the chemical potential $\mu$.
This model has a global $U(1)$ symmetry $\phi \rightarrow e^{i\theta} \phi$  , which is broken for large $\mu$. 

The lattice theory is defined in an ordinary manner. The lattice scalar field
$\phi_n$ lives on a site $n$ of a lattice $\Gamma=\{(n_1,n_2,n_3,n_4)\ \vert n_{\nu}=1,2,\dots ,N_{\nu}\}$ 
with the lattice volume $V=N_1\times N_2\times N_3\times N_4$.
The lattice spacing $a$ is set to $a=1$ without loss of generality.
We choose the periodic boundary condition for the scalar field: $\phi_{n+N_{\nu}{\hat \nu}}=\phi_n$ for $\nu=1,2,3,4$ with ${\hat \nu}$ is the unit vector of the $\nu$-direction. See the $2d$ case~\cite{Kadoh:2019ube} for the other notations. 
Then the corresponding $4d$ lattice action is given by
	\begin{align}
		\label{eq:action}
		S[\phi]=\sum_{n\in\Gamma}	\left\{ (8+m^2)|\phi_n|^2 +\lambda|\phi_n|^4
			-\sum_{\nu=1}^4 \left( e^{\mu\delta_{\nu 4}}\phi_n^\ast\phi_{n+\hat\nu}
				+e^{-\mu\delta_{\nu 4}}\phi_n\phi_{n+\hat\nu}^\ast \right) \right\}. 
	\end{align}
Note that $m$ and $\mu$ in the lattice action are dimensionless ones measured in the lattice unit 
and Eq.~(\ref{eq:cont_action}) is reproduced by taking a naive continuum limit of Eq.~(\ref{eq:action}).
The partition function is defined by
\begin{align}
	Z=\int\mathcal D\phi \, e^{-S[\phi]}
\end{align}
with the path integral measure $\int\mathcal D\phi= \prod_{n \in \Gamma} \int_{-\infty}^{\infty} d {\rm Re}(\phi_n) d {\rm Im}(\phi_n)$.

We employ the polar coordinate $\phi_n=r_n e^{i\pi s_n}$ ($r_n \ge 0$ and $s_n \in [-1,1)$) 
to express the partition function $Z$ as a tensor network.
The action and integral measure are written in
\begin{align}
S[r, s]=\sum_{n\in\Gamma}	\left\{ (8+m^2) r_n^2 +\lambda r_n^4
			-2 \sum_{\nu=1}^4  r_n r_{n+\hat\nu} 
			 \cos \bigg(\pi(s_{n+\hat\nu} - s_n)-i\mu\delta_{\nu 4} \bigg)
			 \right\}
\end{align}
and
\begin{align}
	\int\mathcal D\phi \equiv
	\prod_{n\in\Gamma} \int_0^\infty dr_n r_n \int_{-1}^1 \pi ds_n.
\end{align}
The continuous variables $r_n$ and $s_n$ are discretized by the $K_1$-point Gauss-Laguerre and 
$K_2$-point Gauss-Legendre quadrature rule, respectively: 
$r_{\alpha}$ and $w_{\alpha}$  with $\alpha=1,\ldots, K_1$ denote the $\alpha$th node and weight in the former quadrature and 
$s_{\beta}$ and $u_{\beta}$ with $\beta=1,\ldots,K_2$ are the $\beta$th node and its weight in the latter one.
The partition function is thus discretized as
\begin{align}
	Z(K_1,K_2)= \sum_{ \{\alpha,\beta\} } e^{-S[r,s]}  \prod_{n\in\Gamma} 
		(w_{\alpha_n} e^{r_{\alpha_n}} r_{\alpha_n}) ( \pi u_{\beta_n} )
\end{align}
with
$	\sum_{ \{\alpha,\beta\} }\equiv
	\prod_{n\in\Gamma} \sum_{\alpha_n=1}^{K_1} \sum_{\beta_n=1}^{K_2}.
$

Introducing square matrices 
	\begin{multline}
		M^{[\nu]}_{\alpha\beta,\alpha'\beta'}
		=\sqrt[4]{\pi}
		\sqrt[8]{r_\alpha  r_{\alpha'} w_{\alpha} w_{\alpha'}  u_{\beta}  u_{\beta'}}
		\exp\left( \frac{r_\alpha+r_{\alpha'}}{8} \right)
		\\
		\cdot \exp\left[ \left( 1+\frac{m^2}{8} \right)\left( r_\alpha^2+r_{\alpha'}^2 \right)
		+\frac{\lambda}{8} \left( r_\alpha^4 +r_{\alpha'}^4 \right)
		-2 r_\alpha r_\beta \cos(\pi(s_\beta - s_{\beta'})-i\mu\delta_{\nu 4}) \right]
	\end{multline}
leads to 
\begin{align}
	Z(K_1,K_2)
	=\sum_{ \{\alpha,\beta\} } \prod_{n\in\Gamma} \prod_{\nu=1}^4
	M^{[\nu]}_{\alpha_n\beta_n,\alpha_{n+\hat\nu}\beta_{n+\hat\nu}}.
\end{align}
Note that $M^{[\nu]}$ is the  $(K_1 K_2)\times (K_1 K_2)$ matrix which represents a forward hopping term of the $\nu$-direction.

The singular value decomposition (SVD) is applied to each matrix $M$: 
\begin{align}
	M^{[\nu]}_{\alpha\beta,\alpha'\beta'}
	=\sum_{k=1}^{K_1 K_2}
	U^{[\nu]}_{\alpha\beta,k} \sigma_k^{[\nu]} V^{[\nu]\dagger}_{\alpha'\beta',k}
\label{eq:m_svd}
\end{align}
where $\sigma_k^{[\nu]}$ is the $k$th singular value sorted in the descending order,
and $U^{[\nu]}$ and $V^{[\nu]}$ are the unitary matrices composed of the singular vectors.
Thus, the partition function is represented by a tensor network as
\begin{align}
	\label{eq:Z}
	Z(K_1,K_2)
	=\sum_{x,y,z,t} \prod_{n\in\Gamma}
	T_{x_n y_n z_n t_n x_{n-\hat 1} y_{n-\hat 2} z_{n-\hat 3} t_{n-\hat 4} }
\end{align}
where
\begin{align}
	T_{i_1 i_2 i_3 i_4 j_1 j_2 j_3 j_4 }
	=\sum_{\alpha=1}^{K_1} \sum_{\beta=1}^{K_2} \prod_{\nu=1}^4
	\sqrt{\sigma^{[\nu]}_{i_\nu}\sigma^{[\nu]}_{j_\nu}}
	U^{[\nu]}_{\alpha\beta,i_\nu} V^{[\nu]\dagger}_{\alpha\beta,j_\nu}.
\end{align}
Here tensor indices $x_n,y_n,z_n, t_n, (i_k,j_k)$ run from $1$ to $K_1 K_2$ and the summation $\sum_{x,y,z,t}$ 
is taken for all possible values. 

In actual numerical computations, 
we use a truncated form of Eq.~\eqref{eq:m_svd} up to $D$ $(\le K_1 K_2)$ to reduce the computational cost.
This change is simply understood by redefining $\sum_{x,y,z,t}$ as 
$\sum_{x,y,z,t} \equiv  \prod_{n\in\Gamma} \sum_{x_n=1}^D \sum_{y_n=1}^D \sum_{z_n=1}^D \sum_{t_n=1}^D$. 
The partition function is approximately given with three parameters $D$ and $K_1,K_2$. 
We check the $K_1, K_2$ and $D$ dependences of the results in the next section.

The expectation value of a local operator is also expressed as a tensor network which is 
not homogeneous. 
For instance, the network for $\langle \left| \phi \right|^{2} \rangle$ 
has an impurity tensor which corresponds to the operator insertion in the whole network.
For the particle number density,  two impurity tensors are needed 
as the number operator has a hopping term.

\subsection{ATRG}

The implementation details of the ATRG algorithm is already explained in Ref.~\cite{Akiyama:2019chk}. Here we comment on two points. Firstly, as the partial SVD required in the ATRG, we employ the randomized SVD (RSVD) algorithm, whose accuracy is controlled by the over sampling parameter $p$ and the RSVD iteration number $q$. With the choice of $p\ge 4D$ and $q\ge 2D$, we have confirmed that the numerical results obtained by the ATRG do not depend on these parameters\footnote{For the study of these RSVD parameters in the $4d$ ATRG, see also Ref.~\cite{Oba:2019csk}.}. Secondly, we have slightly modified the original algorithm of ATRG~\cite{Adachi:2019paf}; we avoid taking the square root for the singular values obtained by the RSVD. 
We have checked the improvement of the accuracy with this modification, benchmarking 
with the two-dimensional Ising model.

\section{Numerical results} 
\label{sec:results}

The complex $\phi^4$ theory at finite density is expected to show the Silver Blaze phenomenon where bulk observables are independent of $\mu$ up to some critical point $\mu_c$ in the thermodynamic limit at zero temperature. 
Since this phenomenon is related to the complex phase of the action, 
we examine the $\mu$ dependence of a few observables to test the TRG method.

We choose $m=0.1$ and $\lambda=1$ for the lattice complex $\phi^4$ theory of Eq.~\eqref{eq:action}. These parameters are the same as employed in the $2d$ case of Ref.~\cite{Kadoh:2019ube}.
The partition function of Eq.~\eqref{eq:Z} is evaluated using the ATRG algorithm on a periodic lattice with the volume $V=L^4$ ($L=2^m, m \in \mathbb{Z}$). In the previous section we have introduced two algorithmic parameters. One is the bond dimension of the tensors $D$, which is fixed by keeping the largest $D$ components with the SVD throughout the ATRG algorithm. The other is the polynomial order in the Gauss quadrature method to discretize the complex scalar fields. We set $K=K_{1}=K_{2}$ for simplicity.

\begin{figure}[htbp]
  \centering
  \includegraphics[width=0.8\hsize]{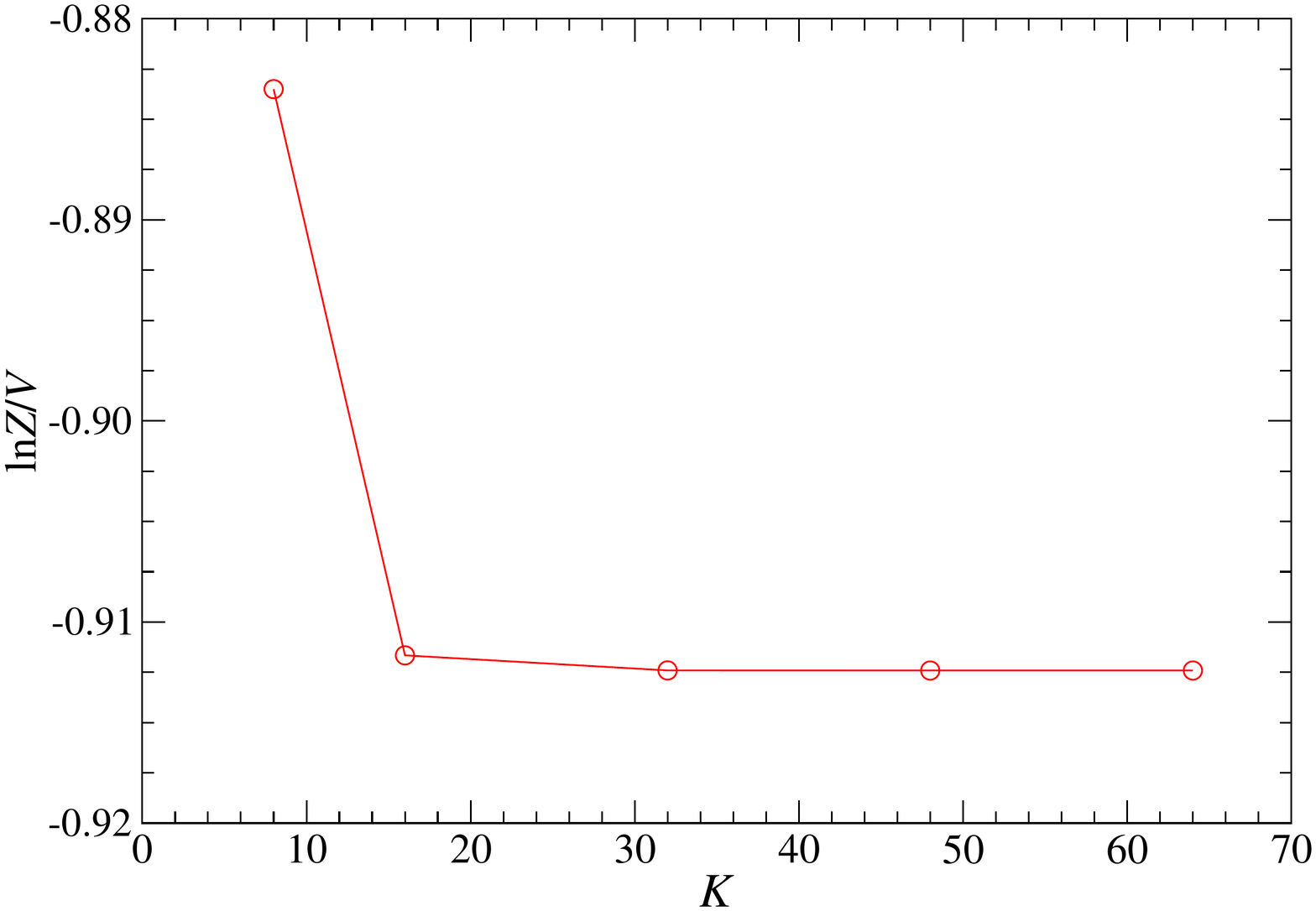}
  \caption{$K$ dependence of thermodynamic potential density with $m^{2}=0.01$, $\lambda=1$, $\mu=0.6$ and $D=45$ on $V=1024^4$.}
  \label{fig:K_vs_F}
\end{figure}

\begin{figure}[htbp]
  \centering
\includegraphics[width=0.8\hsize]{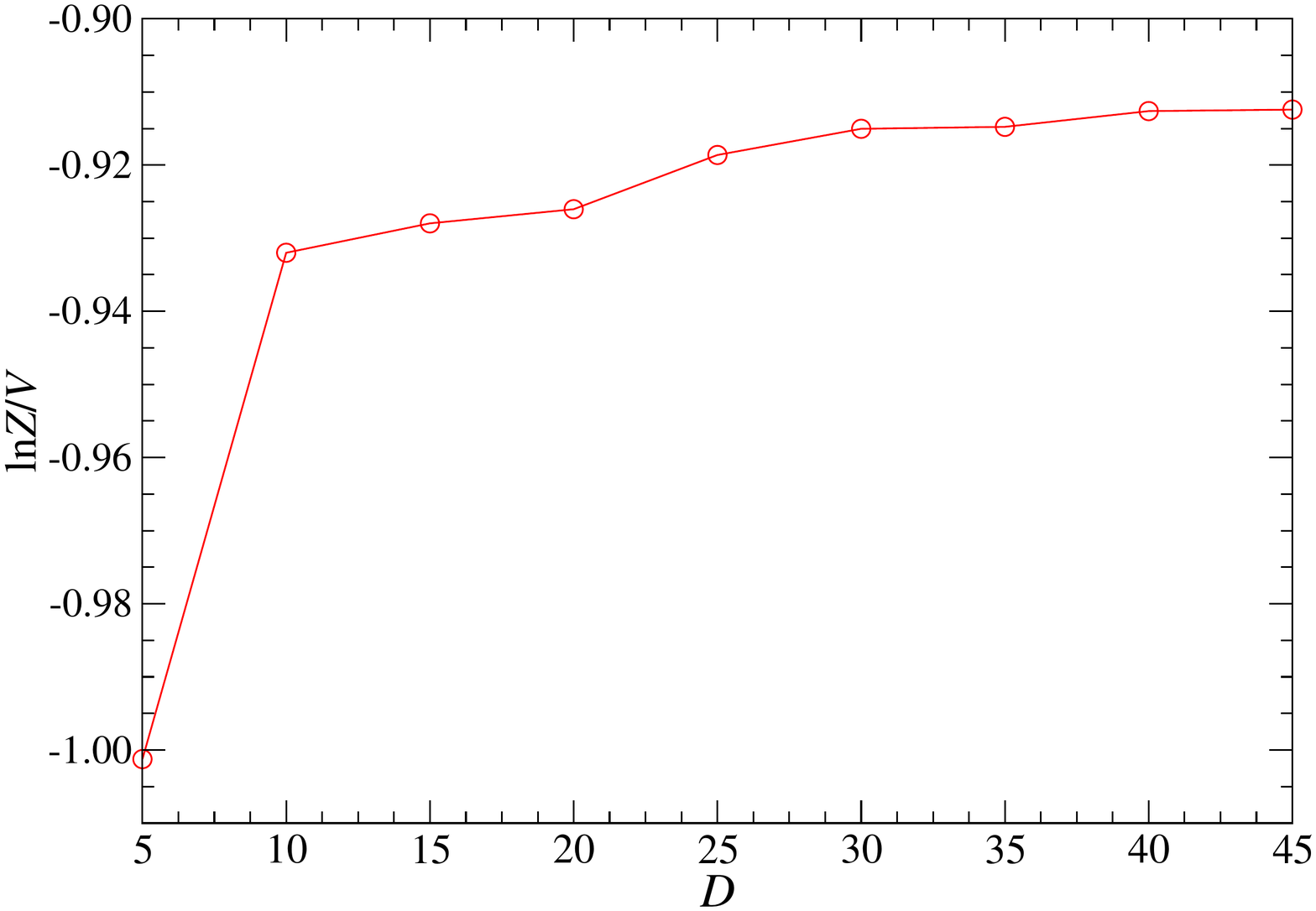}
  \caption{$D$ dependence of thermodynamic potential density with $m^{2}=0.01$, $\lambda=1$, $\mu=0.6$ and $K=64$ on $V=1024^4$.}
  \label{fig:D_vs_F}
\end{figure}

Fig.~\ref{fig:K_vs_F} shows the $K$ dependence  of the thermodynamic potential density $-\Omega=\frac{1}{V}\ln Z$ with $D=45$ on $V=1024^4$ choosing $\mu=0.6$ near the critical chemical potential.
We observe a good convergence behavior for $K$.  
Fig.~\ref{fig:D_vs_F} plots the $D$ dependence of the thermodynamic potential density with $K=64$, 
which seems to be converging around $D\simeq 40$. 
In the following, 
numerical results are presented for $D=45$ and $K=64$ which are large enough in this study.


We first define a phase quenched partition function as
\begin{align}
  \label{eq:zpq}
  Z_{\mathrm{pq}} = \int \mathcal{D} \phi \, e^{-\mathrm{Re}\left( S \right)}, 
\end{align}
dropping the imaginary part of $e^{-S}=e^{-\mathrm{Re}\left( S \right)}e^{i\theta}$ 
with $e^{-\mathrm{Re}(S)}>0$. 
The expectation value of an operator ${\cal O}$ in the phase quenched theory is expressed as  $ \langle {\cal O} \rangle_{\rm pq}$, which is related to $ \langle {\cal O} \rangle$ as 
\begin{align}
\label{ratio_full}
  \langle {\cal O}\rangle
  =
  \frac{\langle {\cal O}e^{i\theta}\rangle_{\rm pq}}{\langle e^{i\theta}\rangle_{\rm pq}}.
\end{align}
In case that the phase factor oscillates frequently in the large $\mu$ region
it is difficult for the Monte Carlo method to evaluate the ratio because of the vanishing contributions from both the numerator and the denominator.
This is the so-called sign problem. 

\begin{figure}[htbp]
  \centering
\includegraphics[width=0.8\hsize]{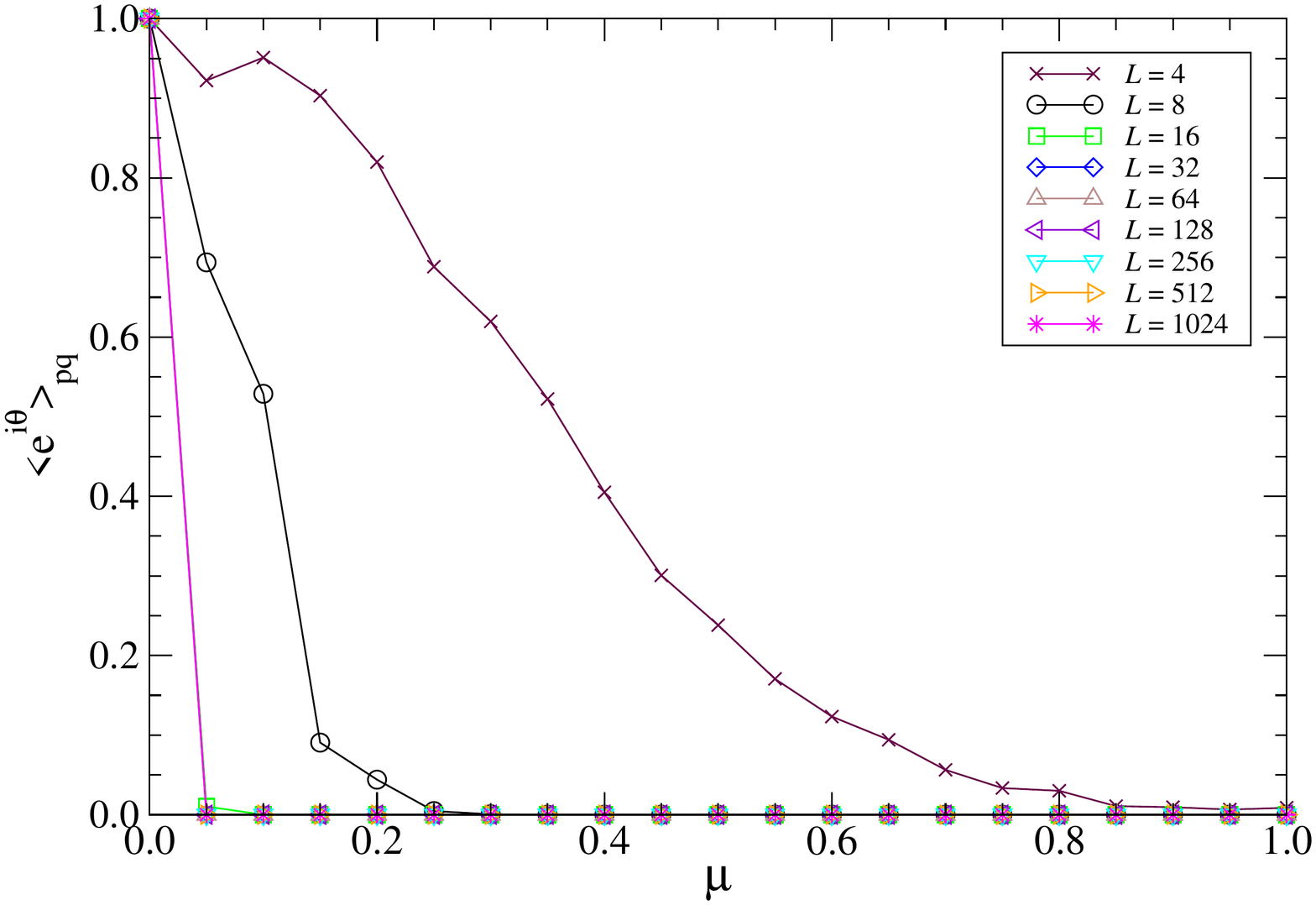}
  \caption{Average phase factor as a function of $\mu$ with $m^{2} = 0.01$, $\lambda=1$, 
  $K=64$, $D=45$. The lattice volume $V$ is varied from $4^4$ to $1024^4$.}
  \label{fig:average_phase}
\end{figure}

In Fig.~\ref{fig:average_phase}, we plot the average phase factor $\langle e^{i\theta}\rangle_{\rm pq}={Z}{/Z_{\mathrm{pq}}}$ 
as a function of $\mu$ varying the lattice volume $V$. 
This quantity measures how severe the sign problem is for given parameters of $\mu$ and $V$. We observe that $\langle e^{i\theta}\rangle_{\rm pq}$ becomes close to zero as either of the volume or the chemical potential increases. 
On the largest volume of $V=1024^4$, which is essentially regarded as the thermodynamic limit at zero temperature, the average phase factor quickly falls off from one at $\mu=0$ to zero for $\mu\gtrsim 0.05$, where a naive Monte Carlo method does not work.

\begin{figure}[htbp]
  \centering
\includegraphics[width=0.8\hsize]{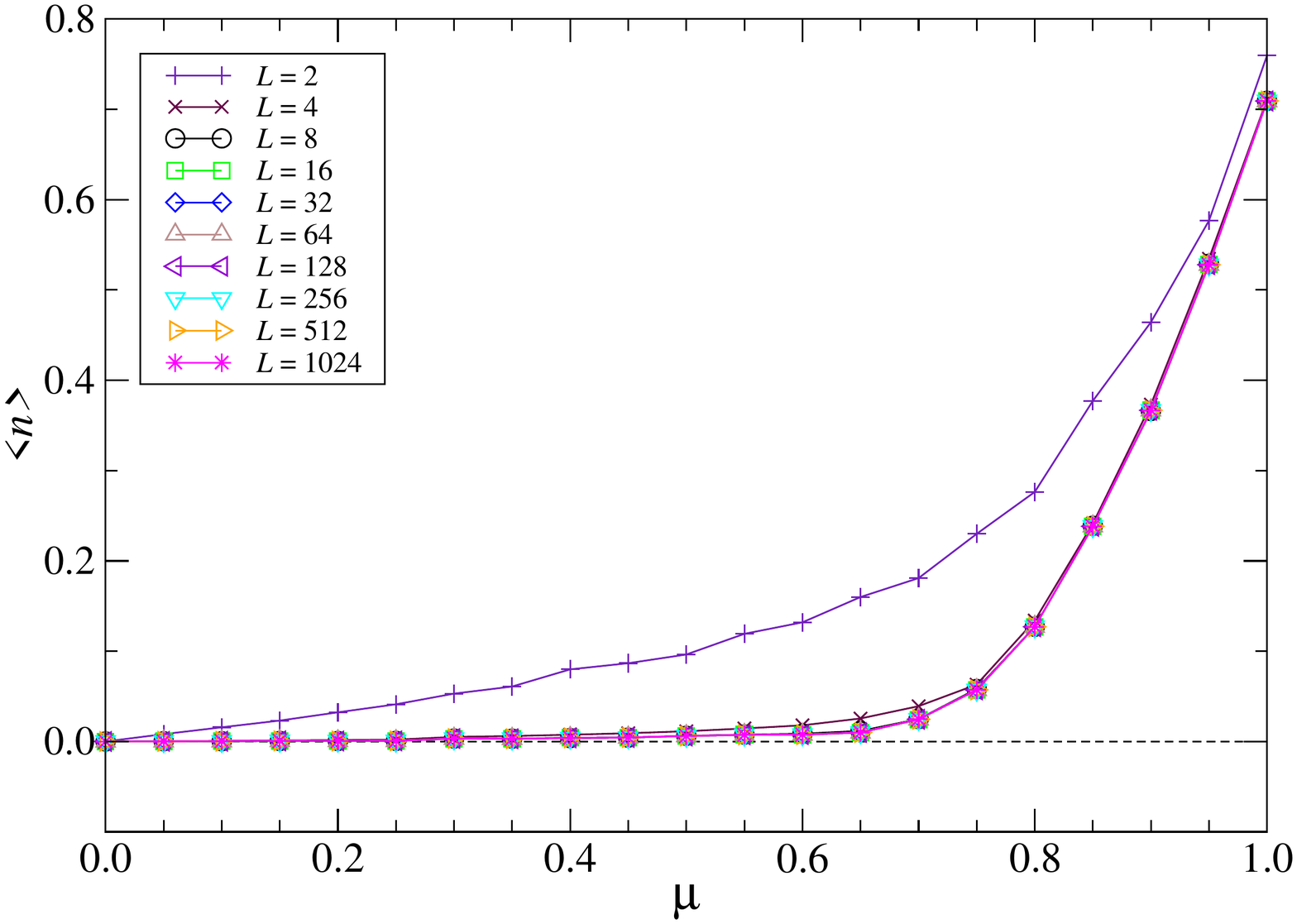}
  \caption{Particle number density as a function of $\mu$ with the lattice volume varied from $2^4$ to $1024^4$.
    The other parameters, $m$, $\lambda$, $K$ and $D$, are the same as those in Fig.~\ref{fig:average_phase}.}
  \label{fig:number_density}
\end{figure}

\begin{figure}[htbp]
  \centering
\includegraphics[width=0.8\hsize]{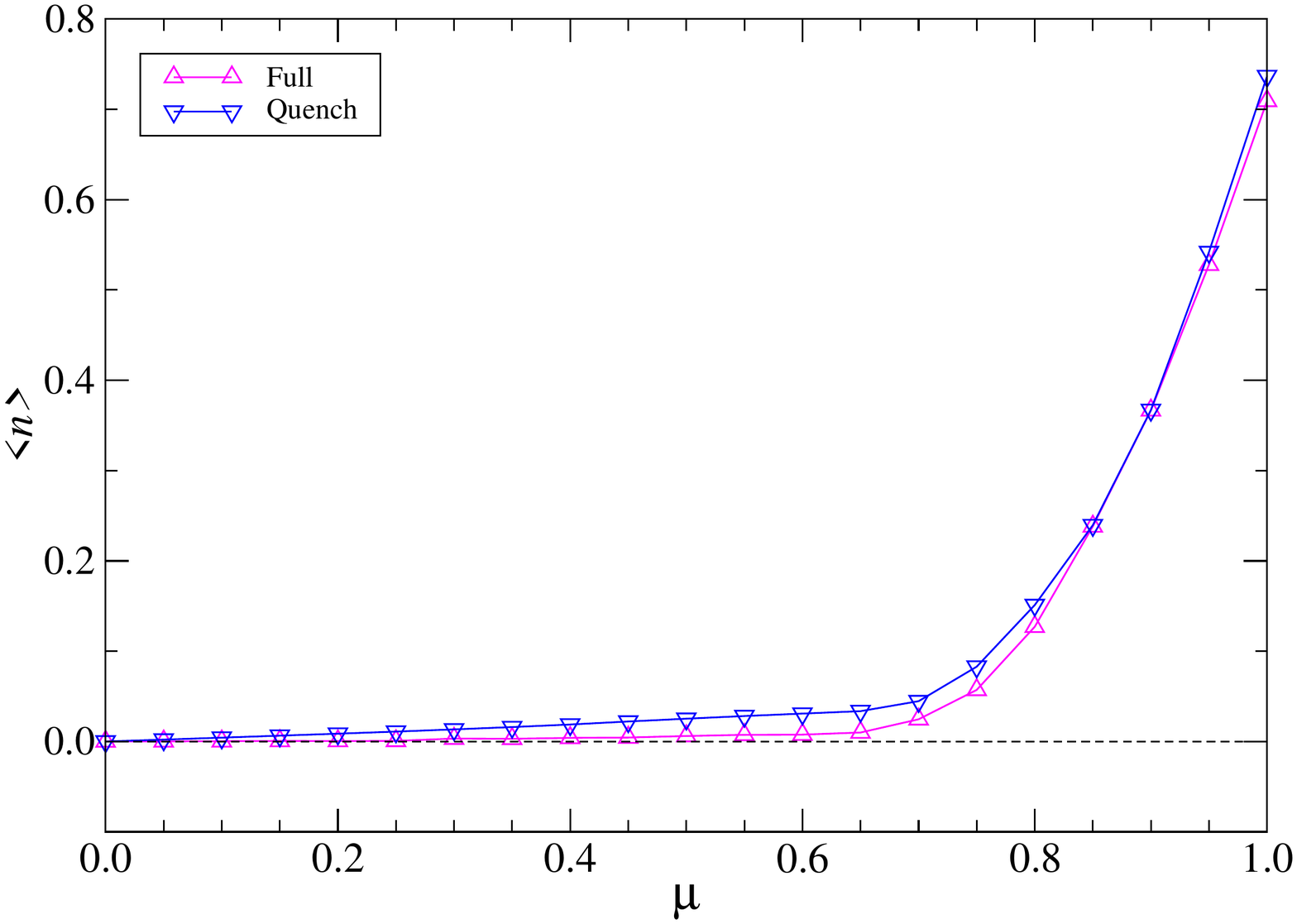}
  \caption{Comparison of $\left< n \right>$ and $\left< n \right>_{\rm pq}$
    with $m^{2} = 0.01$, $\lambda=1$, $K=64$ and $D=45$ on $V=1024^4$.}
  \label{fig:comparison_phase_quench}
\end{figure}

In Fig.~\ref{fig:number_density},  the $\mu$ dependence of the particle number density
\begin{align}
  \left< n \right>=\frac{1}{V}
  \frac{\partial \ln Z}{\partial\mu}
\end{align}
is plotted.
We evaluate this quantity by the ATRG algorithm with impurity tensors \cite{Kadoh:2018tis}. 
On the larger volume 
the Silver Blaze phenomenon becomes manifest: the particle number density stays 
around zero up to $\mu \approx 0.65$ 
and starts to show the rapid increase at $\mu \approx 0.65$ .
Fig.~\ref{fig:comparison_phase_quench} compares $\left< n \right>$ and $\left< n \right>_{\rm pq}$ 
as a function of $\mu$ on the lattice of $V=1024^4$. 
The latter shows the monotonic increase once the finite chemical potential is turned on. 
It is confirmed that the Silver Blaze phenomenon is attributed to the imaginary part of the action.

\begin{figure}[htbp]
  \centering
\includegraphics[width=0.8\hsize]{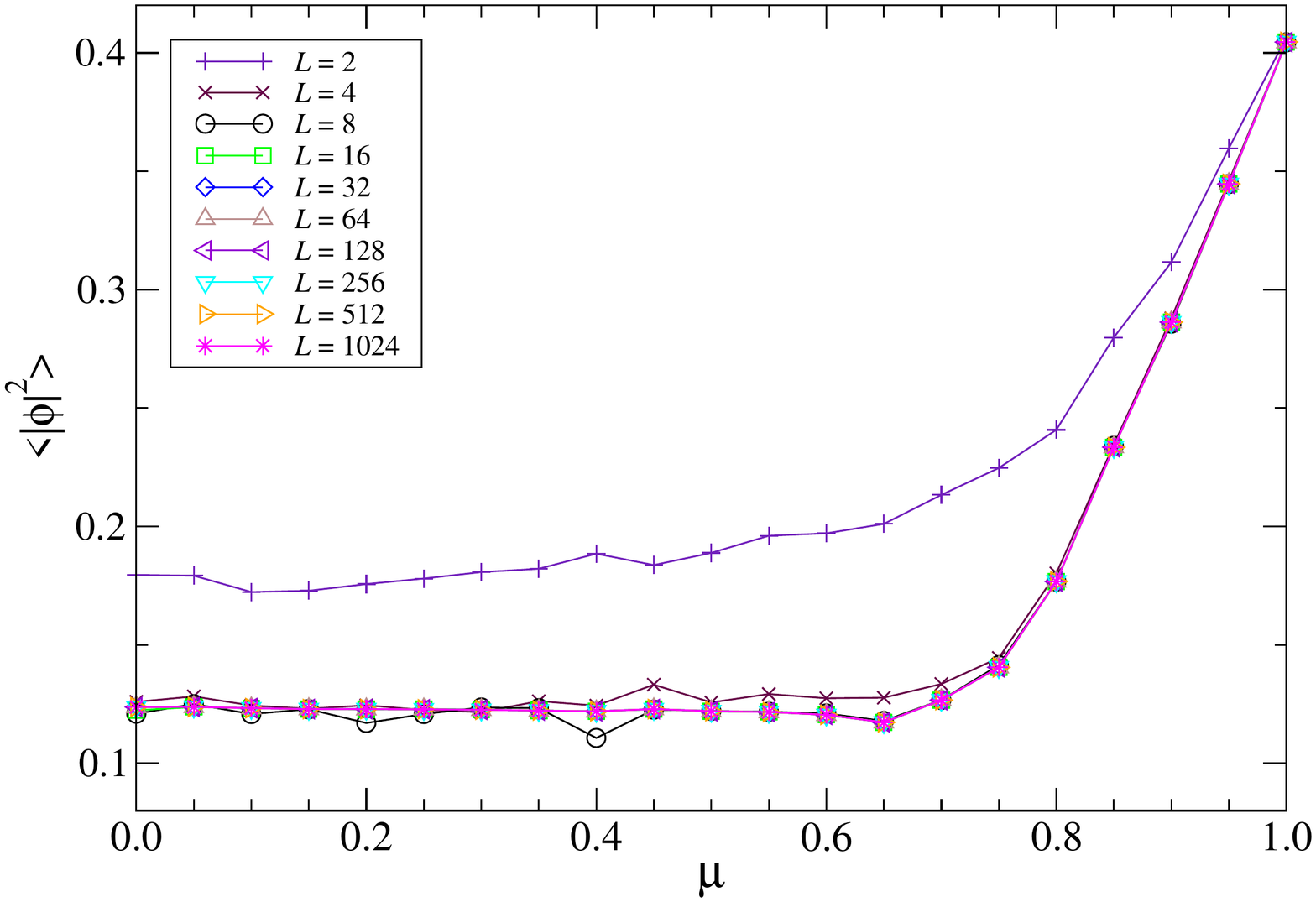}
  \caption{ $\langle \left| \phi \right|^{2} \rangle$ as a function of $\mu$ with the same parameter set of $(m,\lambda,K,D,V)$  as in Fig.~\ref{fig:number_density}.}
  \label{fig:phi_squared}
\end{figure}

Fig.~\ref{fig:phi_squared} plots $\langle \left| \phi \right|^{2} \rangle$ as a function of $\mu$ with the same parameter set of $(m,\lambda,K,D,V)$ as in Fig.~\ref{fig:number_density}, 
which is  also evaluated with the impurity tensor method.
The $\mu$ dependence of $\langle \left| \phi \right|^{2} \rangle$ is quite similar to that of $\left< n \right>$ in Fig.~\ref{fig:number_density}: $\langle \left| \phi \right|^{2} \rangle$ seems independent of $\mu$ up to $\mu\approx 0.65$ and shows the rapid increase beyond it.  The value of the critical chemical potential $\mu_c$ should be compared with a mean field estimate, which is given by $4\sinh^2(\mu_c^{\rm MF}/2)=m^2+4\lambda \langle \left| \phi \right|^{2} \rangle$~\cite{Aarts:2009hn}. Using the measured value of $\langle \left| \phi \right|^{2} \rangle\approx 0.125 $ over $0\le \mu \le 0.65$ in Fig.~\ref{fig:phi_squared} we obtain $\mu_c^{\rm MF}\approx 0.70$. 

\section{Summary and outlook} 
\label{sec:summary}

We have applied the ATRG algorithm to study the $4d$ complex $\phi^4$ theory at finite density. This is the first attempt to analyze a $4d$ quantum field theory with the TRG approach. The Silver Blaze phenomenon is clearly observed for the particle number density and $\langle \left| \phi \right|^{2} \rangle$ on the extremely large lattice of $V=1024^4$ which is essentially in the thermodynamic limit at zero temperature. This successful study encourages us to extend the analysis to other $4d$ quantum field theories.  

\begin{acknowledgments}
Numerical calculation for the present work was carried out with the Oakforest-PACS (OFP) and the Cygnus computers under the Interdisciplinary Computational Science Program of Center for Computational Sciences, University of Tsukuba.
This work is supported by the Ministry of Education, Culture, Sports, Science and Technology (MEXT) as ``Exploratory Challenge on Post-K computer (Frontiers of Basic Science: Challenging the Limits)'' and JSPS KAKENHI Grant JP19K03853, 20H00148.
\end{acknowledgments}

\bibliographystyle{JHEP}

\bibliography{algorithm,discrete,grassmann,continuous,gauge,cphi4}

\end{document}